\documentstyle{mn}

\newif\ifAMStwofonts

\ifoldfss
  \ifCUPmtlplainloaded \else
    \NewTextAlphabet{textbfit} {cmbxti10} {}
    \NewTextAlphabet{textbfss} {cmssbx10} {}
    \NewMathAlphabet{mathbfit} {cmbxti10} {} 
    \NewMathAlphabet{mathbfss} {cmssbx10} {} 
  \fi
  \ifAMStwofonts
    \ifCUPmtlplainloaded \else
      \NewSymbolFont{upmath} {eurm10}
      \NewSymbolFont{AMSa} {msam10}
      \NewMathSymbol{\upi}     {0}{upmath}{19}
      \NewMathSymbol{\umu}     {0}{upmath}{16}
      \NewMathSymbol{\upartial}{0}{upmath}{40}
      \NewMathSymbol{\leqslant}{3}{AMSa}{36}
      \NewMathSymbol{\geqslant}{3}{AMSa}{3E}

    \fi
  \fi
\fi 

\ifnfssone
  \newmathalphabet{\mathit}
  \addtoversion{normal}{\mathit}{cmr}{m}{it}
  \addtoversion{bold}{\mathit}{cmr}{bx}{it}
  \newmathalphabet{\mathbfit} 
  \addtoversion{normal}{\mathbfit}{cmr}{bx}{it}
  \addtoversion{bold}{\mathbfit}{cmr}{bx}{it}
  \newmathalphabet{\mathbfss} 
  \addtoversion{normal}{\mathbfss}{cmss}{bx}{n}
  \addtoversion{bold}{\mathbfss}{cmss}{bx}{n}
  \ifAMStwofonts
    \ifCUPmtlplainloaded \else
      \UseAMStwoboldmath
      \makeatletter
      \new@mathgroup\upmath@group
      \define@mathgroup\mv@normal\upmath@group{eur}{m}{n}
      \define@mathgroup\mv@bold\upmath@group{eur}{b}{n}
      \edef\UPM{\hexnumber\upmath@group}
      \new@mathgroup\amsa@group
      \define@mathgroup\mv@normal\amsa@group{msa}{m}{n}
      \define@mathgroup\mv@bold\amsa@group{msa}{m}{n}
      \edef\AMSa{\hexnumber\amsa@group}
      \makeatother
      \mathchardef\upi="0\UPM19
      \mathchardef\umu="0\UPM16
      \mathchardef\upartial="0\UPM40
      \mathchardef\leqslant="3\AMSa36
      \mathchardef\geqslant="3\AMSa3E
    \fi
  \fi
\fi 

\ifnfsstwo
  \DeclareMathAlphabet{\mathbfit}{OT1}{cmr}{bx}{it}
  \SetMathAlphabet\mathbfit{bold}{OT1}{cmr}{bx}{it}
  \DeclareMathAlphabet{\mathbfss}{OT1}{cmss}{bx}{n}
  \SetMathAlphabet\mathbfss{bold}{OT1}{cmss}{bx}{n}
  \ifAMStwofonts
    \ifCUPmtlplainloaded \else
      \DeclareSymbolFont{UPM}{U}{eur}{m}{n}
      \SetSymbolFont{UPM}{bold}{U}{eur}{b}{n}
      \DeclareSymbolFont{AMSa}{U}{msa}{m}{n}
      \DeclareMathSymbol{\upi}{0}{UPM}{"19}
      \DeclareMathSymbol{\umu}{0}{UPM}{"16}
      \DeclareMathSymbol{\upartial}{0}{UPM}{"40}
      \DeclareMathSymbol{\leqslant}{3}{AMSa}{"36}
      \DeclareMathSymbol{\geqslant}{3}{AMSa}{"3E}
    \fi
  \fi
\fi 

\ifCUPmtlplainloaded \else
  \ifAMStwofonts \else 
    \def\upi{\pi}
    \def\umu{\mu}
    \def\upartial{\partial}
  \fi
\fi

\title{Hipparcos absolute magnitudes for metal rich K giants and the
calibration of DDO photometry}

\author[E. H\o g and C. Flynn]{Erik H\o g$^{1,3}$ and Chris Flynn$^{2,4}$\\ 
$^1$Copenhagen University Observatory, Juliane Maries Vej 30, DK 2100
Copenhagen\\
$^2$Tuorla Observatory, Piikki\"o, FIN-21500, Finland\\
$^3$erik@astro.ku.dk \\ $^4$cflynn@astro.utu.fi }

\date{Accepted, Received ; in original form }

\begin{document}
\maketitle

\begin{abstract}

  Parallaxes for 581 bright K giants have been determined using the Hipparcos
satellite. We combine the trigonometric parallaxes with ground based
photometric data to determine the K giant absolute magnitudes. For all these
giants, absolute magnitude estimates can also be made using the intermediate
band photometric DDO system (Janes 1975, 1979).  We compare the DDO absolute
magnitudes with the very accurate Hipparcos absolute magnitudes, finding
various systematic offsets in the DDO system.  These systematic effects can be
corrected, and we provide a new calibration of the DDO system allowing absolute
magnitude to be determined with an accuracy of 0.35 mag in the range $2 > M_V >
-1$. The new calibration performs well when tested on K giants with DDO
photometry in a selection of low reddening open-clusters with well-measured
distance moduli.


\end{abstract}

\section{Introduction}

  K giants are bright and ubiquitious, occuring in a wide range of Galactic
populations, and are convenient tracer objects for examining the structure and
kinematics of the Galaxy.  The chief difficulty with these objects has been the
uncertainty in their absolute magnitudes, which arises from the fact that stars
on the giant branch form from a wide range of mass and age, as well as the giant
branch being rather steep as a function of colour.  K giants span a broad range
of absolute magnitude $M_V$ from about $2 < M_V < -3$.

  In some studies accurate distances to the tracer objects are required such as
in Bahcall, Flynn and Gould (1991) and Flynn and Fuchs (1994), who used the
kinematics of K giants to constrain the amount of dark matter present in the
Galactic disc. In these studies the absolute magnitudes and hence distances of
the giants were estimated using David Dunlop Observatory (DDO) photometry.
This is an intermediate band photometric system of six filters, four of which
can be used to estimate physical parameters for late type giants (McClure
1976). The four filters are called 41, 42, 45 and 48 and are at the wavelengths
4166, 4257, 4517 and 4886 \AA~and have passbands of 83, 73, 76 and 186
\AA~respectively. Three colours, C4142, C4548 and C4245 are formed from these
four filters.  C4245 is primarily sensitive to effective temperature, C4548 to
luminosity and C4142 can be used in combination with the other two colours to
estimate stellar metallicity, [Fe/H]. Janes (1979) describes the measurement of
absolute magnitude and abundance using these filters.  His absolute magnitude
scale was based on distances determined by means of the Wilson-Bappu effect,
and was later adjusted slightly when DDO photometry had been obtained of K
giants in open clusters.  Janes' calibration applies to metal rich disc stars
([Fe/H] $>-1$), although the system has been extended to metal weak populations
(e.g. Norris, Bessell and Pickles 1985, Morrison, Flynn and Freeman 1990, and
Claria et.al. 1994).  In this paper we are concerned with metal rich K giants
only, [Fe/H]$>-0.5$.

  The {\it European Space Agency's} Hipparcos satellite has observed all bright
(apparent $V<8.0$) K giants, which are included in the all sky part of the
Hipparcos Input Catalogue (HIC), so that very high accuracy trigonometric
parallaxes are now available in the Hipparcos Catalogue (ESA 1997). Before
Hipparcos, only a handful of giants had accurate parallax measurements, whereas
Hipparcos has now measured accurate parallaxes for around 600 giants. In this
paper, we use the parallaxes of local K giants from Hipparcos to check the DDO
system's absolute magnitude calibration. Our sample is described in section 2.
A number of systematic offsets are found in the DDO absolute magnitudes,
particularly for the redder ($B-V>1.2$) giants and around the clump giants.  In
section 3 we develop a new calibration of the DDO system, tied to the Hipparcos
results. In addition to removing the systematic errors in the old system, the
new calibration is a good deal simpler to use.  Absolute magnitudes of K-giants
can thus be determined photometrically over the range $2 < M_V < -1$ with an
accuracy of 0.34 magnitudes. In section 4 the new calibration is found to be
satisfactory when checked with K giants in old open clusters for which DDO
photometry is available in the literature. The calibration developed here can
be applied in a wide range of Galactic structure studies, one of which is the
measurement of the amount of dark matter in the Galactic disc.  We draw our
conclusions in section 5.

\section{K giant sample and analysis}

  Our specific purpose was to calibrate the distance scale for metal rich K
giants, [Fe/H] $> -0.5$, since it is these giants which are of particular
interest in the examination of the mass density of the local disc. Accordingly,
we selected all G and K giants in the Bright Star Catalog brighter than V=6.5
and for which DDO photometry is available from Mclure and Forrester (1981). We
selected stars in the colour range $0.85<$ C4245 $<1.15$, which corresponds
closely to the $B-V$ range $1.0 < B-V < 1.35$ (Flynn and Freeman 1993) and
isolates (approximately) G8 to K3 giants.  All the choosen objects are well
within the magnitude limit of Hipparcos (which is complete over the whole sky
to $V=7.3$), so all were measured by the satellite. The parallaxes $\pi$
obtained by the satellite have errors $\sigma_\pi$ of typically 0.8 mas (mas =
milliarcsec).  There were 676 giants in this basic sample.  We obtained
Hipparcos data for the sample in the first PI release, in August 1996. For each
object, we used the Hipparcos measurement of the parallax $\pi$, the parallax
error $\sigma_\pi$ and the apparent $V$ magnitude to calculate the absolute
magnitude $M_V$ and error in the absolute magnitude $\delta M_V$. In Figure 1
we show the distribution of parallaxes $\pi$, the distribution of the relative
parallax error $\sigma_\pi/\pi$ and the distribution of absolute magnitude
error $\delta M_V$. These figures demonstrate the very high quality of the
Hipparcos data. A typical giant in this sample has a parallax of 10 mas, a
relative parallax error of 8\% and an absolute magnitude error of 0.15 mag. We
used the apparent $V$ magnitudes in the Hipparcos Input Catalog, which are
accurate to 0.01 mag., and hence the parallax error completely dominates the
absolute magnitude error.

\begin{figure} 
\input epsf
\centering
\leavevmode
\epsfxsize=0.9
\columnwidth
\epsfbox{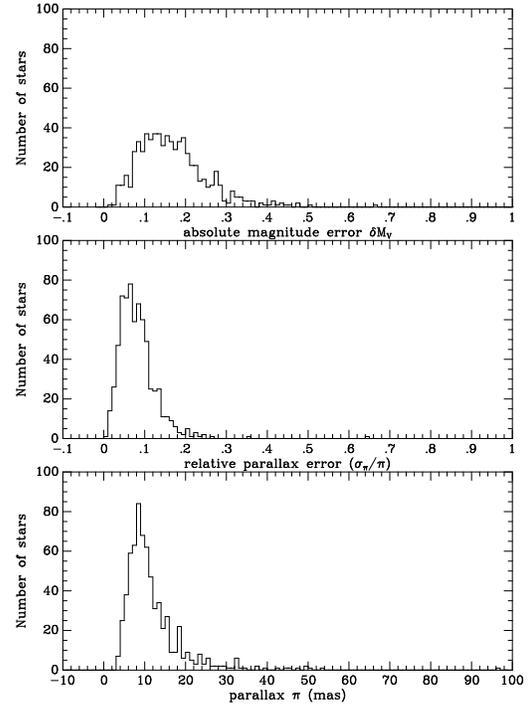}
\caption{The Hipparcos results for the basic sample of 676 metal rich
K giants. Lower panel: distribution of parallax $\pi$ in mas
(milliarcsec).  A typical giant has a parallax of 10 mas.  Middle
panel: distribution of the relative standard error in the parallaxes
$\sigma_\pi/\pi$. Upper panel: distribution of the absolute magnitude
standard error $\delta M_V$ for the giants.  A typical giant has a
standard error of 0.15 mag.}
\end{figure}

  Most of these stars are closer to the Sun than 120 pc so that the reddening
to each star is expected to be small. We checked this by calculating the
reddening to each star using the method of Janes (1975) from the DDO photometry
and $B-V$ colour. The mean reddening of the sample was found to be
$E(B-V)=0.01$ with a scatter of $0.03$ dex.  A histogram of the reddening
values is shown in Figure 2. The scatter is consistent with the accuaracy of
the method of 0.03 dex (Janes 1975). For a handful of stars a small amount of
reddening was indicated, and so stars with $(E(B-V) > 0.05$ were removed from
the sample. For the remaining stars the mean reddening reduces to $E(B-V)=0.00$
so no reddening corrections were deemed necessary.

  Finally we confined ourselves to stars with a relative parallax error of less
than 0.15 (or an absolute magnitude error of less than 0.32 mag --- see Figure
1). This reduces the sample to 581 giants and places a distance limit on the
sample of approximately 120 pc.

  The colour magnitude diagram (CMD) of our selected giants is shown in Figure
3(a), where $B-V$ colour from the Hipparcos Catalogue is plotted versus the
absolute mangitude, $M_V(\mathrm{Hipp})$, determined from the Hipparcos
parallaxes and the V magnitude in the Hipparcos catalogue. The prominent
features are the steeply rising giant branch and the clump stars near
$M_V=0.8$. The typical error in absolute magnitude is 0.15 mag.

  For each giant an absolute magnitude $M_V(DDO)$ was estimated using Janes
(1979) calibration.  Figure 3(b) shows the CMD using this method for the same
giants as Figure 3(a).  Hipparcos clearly reveals that there are two main
problems in the DDO calibration.  Firstly, $M_V(DDO)$ for the redder giants
($B-V > 1.2$) is underestimated (i.e. Hipparcos reveals that they are brighter
than previously thought). Secondly, the position of the clump stars (the core
He burning giants in the field equivalent of the Horizontal Branch) is
estimated too bright (by approx 0.3 mag) with the DDO system. Note that the
clump stars show up as a much sharper feature in the Hipparcos diagram because
the typical errors in the absolute magnitudes are much smaller.

\begin{figure} 
\input epsf
\centering
\leavevmode
\epsfxsize=0.9
\columnwidth
\epsfbox{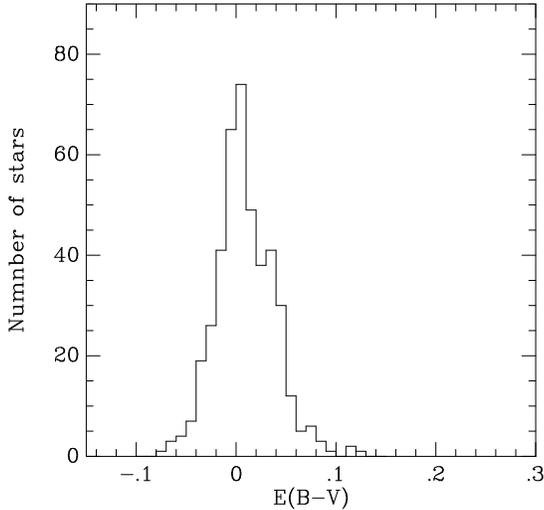}
\vskip -2.5 cm
\caption{Histogram of the individual reddenings of the metal rich
giants in the sample.  A small number of stars with $E(B-V) > 0.05$
were removed from the sample.  The remaining stars are consistent with
zero reddening, as expected in this nearby sample.}
\end{figure}

\begin{figure} 
\input epsf
\centering
\leavevmode
\epsfxsize=0.9
\columnwidth
\epsfbox{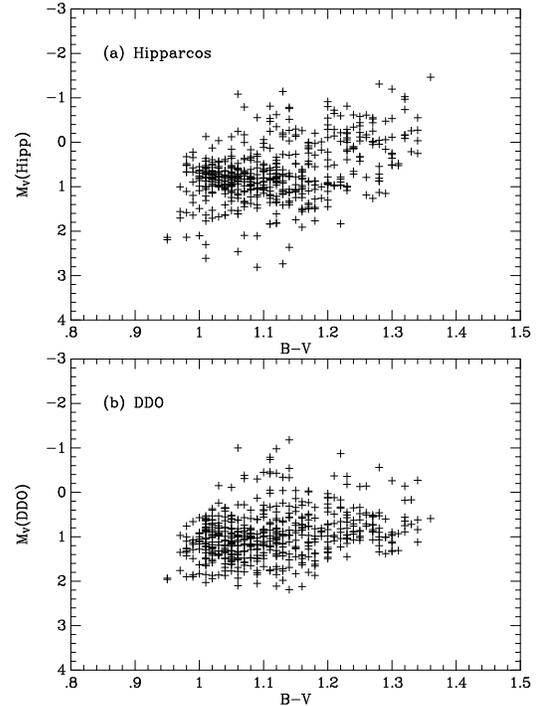}
\caption{Colour magnitude diagrams of the giant branch. The upper
panel comes from the Hipparcos data. Typical error in the absolute
magnitude $M_V(\mathrm{Hipp})$ is only 0.15 mag, approximately the
symbol size.  The steeply rising giant branch and the clump are the
salient features of the plot. With such accurate magnitudes, the clump
appears sharp at an absolute magnitude of $M_V = 0.8$. In the lower
panel, the absolute magnitudes for the exact same giants, using the
old DDO photometric system. These absolute magnitudes have lower accuracy,
0.3 to 0.5 mag. depending on absolute magnitude.  While the clump is
clearly seen, the feature is not as sharp as in the Hipparcos
data. Importantly, the clump is estimated about 0.3 mag too faint by
the DDO method. Furthermore, the absolute magnitudes of the redder
giants ($B-V > 1.2$) are estimated by DDO as considerably too faint;
Hipparcos shows that they are actually about 1 magnitude brighter than
DDO predicts.}
\end{figure}

\subsection{Malmquist bias and Lutz-Kelker corrections}

  In any sample of parallaxes which is magnitude limited like this one, there
are potentially biases in the derived absolute magnitudes due to Malmquist bias
and the Lutz-Kelker (1973) effect.

  The simplest method to determine the size of these biases was to carry out a
Monte-Carlo simulation of our selection procedure on simulated catalogs of
stars. The results show that for this particular sample such biases are smaller
than 0.01 mag. and can be safely ignored. This is primarily due to the high
accuracy of the Hipparcos parallaxes and the very small number of stars
rejected because of a large relative parallax error.

  We carried out simulations in which we distributed 10000 stars at random with
uniform density in a sphere around the Sun to a radius of 400 pc, this being
beyond the most distant stars in the observed sample. The stars were assumed to
have the same absolute magnitude $M_V$. We know the true distance and true
parallax $\pi$ for each star, and we simulate the measured parallax $\pi({\rm
obs})$ for each star from $\pi$ by adding an error drawn from a Gaussian
distribution of dispersion $\sigma_\pi = 0.8$ mas.

  The main selection criteria of the real sample of stars are that $V < 6.5$
and that the relative parallax error $\sigma_\pi/\pi < 0.15$.  For all stars
meeting these selection criteria in the simulation, we compare the true
absolute magnitude $M_V$ and the absolute magnitude $M_V({\rm obs})$ we would
derive from the observed parallax $\pi({\rm obs})$. We find that, over the
range in absolute magnitude of interest, $2 \ga M_V \ga -1$, the mean
difference between true and measured absolute magnitudes is less than 0.01 mag.

  We assumed the stars are uniformly distributed in space, which is a
reasonable first approximation since most of the stars in the sample are closer
than 120 pc. The stars will actually have some vertical density falloff, and we
tested the effect of an exponential falloff where the density $\rho$ of the
stars falls with height $z$ above the Galactic plane like $\rho(z) = \rho(0)
e^{-z/h}$, with the scale height $h$ ranging between 100 and 300 pc, typical
values for the Galactic disk. As expected, no systematic differences between
true and measured absolute magnitude were found. We conclude from our
simulations that our estimates of the stellar absolute magnitudes based on
$1/\pi$ require no corrections for Malmquist or Lutz-Kelker bias.

\section{DDO2 : An metal rich K-giant absolute magnitude estimator}

\subsection{The initial calibration}  

The DDO absolute magnitudes clearly show several systematic differences
relative to the Hipparcos data. We decided to recalibrate the DDO system's
absolute magnitude estimation system from scratch, rather than figure out a
correction scheme. We term this new method DDO2. The DDO absolute magnitudes
are derived from C4245 and C4548 colours, with some small corrections for the
metallicity of the star (but note that [Fe/H] is derived from all three colours
C4245, C4548 and C4142).  We made a file containing for each star: the
Hipparcos measured absolute magnitude, the three DDO colours, C4245, C4548,
C4142, the broadband $B-V$ colour and the abundance [Fe/H] (derived in the
usual way from DDO colours using Janes 1979). We then searched amongst these
parameters for a simple fit to $M_V(\mathrm{Hipp})$ using a surface fitting
package (TableCurve 3-D, from Jandel Scientific), which allows one to inspect
visually a wide range of fitting surface types for two dependent parameters to
a third parameter. C4245 and C4548 emerged quickly as the primary dependent
parameters, and the absolute magnitudes could be fit in terms of these two
colours by the following quadratic polynomial:

\begin{equation}
 M_V(1) = a + bx + cy + dx^2 + ey^2 + fxy.
\end{equation}

\noindent where $x = $ C4245 and $y = $ C4548. The coefficients of the best fit
are shown in Table 1. The scatter in the fit is 0.34 magnitudes. We tried a
number of higher order polynomials, but could obtain no significant improvement
on this fit. The quadratic gave a clearly better fit than a simple plane
(i.e. $d = e = f = 0$).  The fitted absolute magnitudes $M_V$ are shown as a
function of the $M_V(\mathrm{Hipp})$ in Figure 3.

  We checked our overall fit for possible systematic effects as follows.
Firstly, we plotted the residuals as a function of other known observational
properties of the giants: C4142, [Fe/H], $B-V$ colour and right ascension
$\alpha$ and declination $\delta$ (the latter two in case of any systematic
effects around the sky). A small correction for metallicity emerged:

\begin{figure} 
\input epsf
\centering
\leavevmode
\epsfxsize=0.9
\columnwidth
\epsfbox{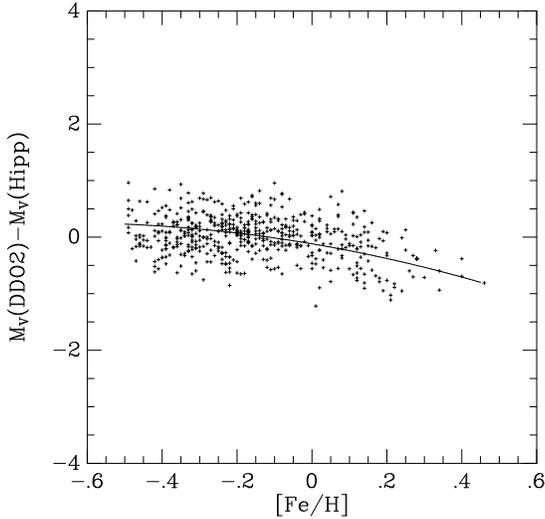}
\vskip -2 cm
\caption{Residual abundance dependence of the fit to the Hipparcos
absolute magnitudes. The line shows the adopted correction (Equation
2) for this residual dependence with [Fe/H].}
\end{figure}

\begin{equation}
 \Delta M_V(1) = a_0 + a_1\times{\mathrm{[Fe/H]}} +
  a_2\times{\mathrm{[Fe/H]}}^2
\end{equation}

The coefficients $(a_0, a_1, a_2)$ are shown in Table 1 and the correction is
illustrated in Figure 4. 

\subsection{The final DDO2 calibration} 

  Equations (1) and (2) lead to an absolute magnitude estimator for which no
residuals remain as a function of colour or metallicity for the 581 giants as a
whole.  However, since the evolutionary state of the giants varies
significantly over the colour-magnitude diagram (in particular we have both
clump and first ascent giants in the sample), it remained possible that
interplay between colour and metallicity could lead to measurable systematic
offsets in particular regions of the colour-magnitude diagram. For example in
the clump region, metal poor giants tend to be bluer and the metal rich ones
redder, while on the Giant Branch, metal poor giants are bluer and brighter
than metal rich giants.

  We investigated the possible systematic effects as follows: we divided the
colour-magnitude diagram into a grid of boxes 0.1 mag. wide in colour and 0.5
to 1.0 mag. wide in $M_V$. There were typically 30 giants in each box with
which to check for systematic differences in the estimated and true (Hipparcos)
absolute magnitude as functionos of colour or abundance.  No systematics were
found in the areas dominated by the first ascent giants (in particular the
region $B-V > 1.15$). A clear trend was found in the horizontal branch ($B-V<
1.15$) region. Hence, in the region $B-V < 1.15$ there is a correction to the
absolute magnitudes of

\begin{equation}
 \Delta M_V(2) = b_0 + b_1\times{\mathrm{[Fe/H]}} ~~~~~\mathrm{for}~~ B-V < 1.15 
\end{equation}

\noindent This leads to our DDO2 absolute magnitude estimator :

\begin{equation}
 M_V({\mathrm{DDO2}}) = M_V(1) + \Delta M_V(1) + \Delta M_V(2).
\end{equation}

  Figure 5 shows the CMD of the giants using this transformation
$M_V(\mathrm{DDO2})$, including the metallicity corrections. Comparing this
with the Hipparcos measured CMD (Figure 3a) one sees that with the new system
the positions of the clump stars and the redder giants, as well as the CMD
magnitude in general, are more satisfactory than with the old DDO system. In
Figure 6 we show the residuals between $M_V(\mathrm{Hipp})$ and
$M_V(\mathrm{DDO2})$ as functions of $B-V$, [Fe/H] and $M_V(\mathrm{DDO2})$, to
illustrate that no significant reidual trends are apparent.

\begin{table}
\caption{Best fit coefficients of the surface fit to $M_V$ and the
residual [Fe/H] dependence (Equations 1, 2 and 3)}
\begin{center}
\begin{tabular}{cr}
\hline
     $ a$ & $   -3.5185187$\\  
     $ b$ & $   31.6767174$\\  
     $ c$ & $    5.3719567$\\  
     $ d$ & $  -28.1707768$\\  
     $ e$ & $  -27.7136769$\\  
     $ f$ & $   29.6306218$\\
     $ a_0$& $ 0.058910365$\\
     $ a_1$& $ 1.12501061$\\
     $ a_2$& $ 0.85962826$\\
     $ b_0$& $ 0.051$\\
     $ b_1$& $ 0.897$\\
\hline
\end{tabular}
\end{center}  
\end{table}

  In summary, the new calibration of DDO photometry can be used to derive
absolute magnitudes for metal rich ([Fe/H] $> -0.5$) K giants with an accuracy
of 0.34 dex over the absoute magnitude range $2 > M_V > -1$ and in the colour
range $0.85 < $C4245$ < 1.15$.

\begin{figure} 
\input epsf
\centering
\leavevmode
\epsfxsize=0.9
\columnwidth
\epsfbox{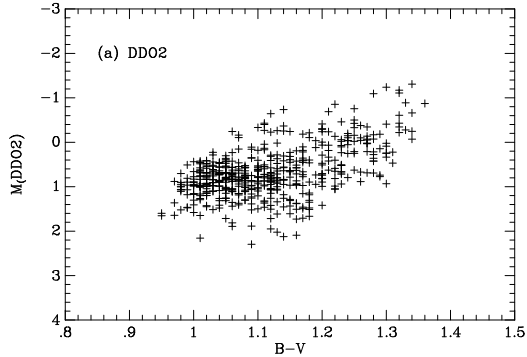}
\vskip -3.5 cm
\caption{Colour magnitude diagram of the sample giants using the new
DDO2 calibration to calculate absolute magnitude from DDO photometry.
This diagram can be compared directly with the Hipparcos colour
magnitude diagram Figure 3(a). Generally, the new calibration produces
a better resemblance to the Hipparcos colour magnitude diagram, and in
particular the new calibration places the clump stars and the redder
giants ($B-V> 1.2)$ close to their true positions.}
\end{figure}

\begin{figure} 
\input epsf
\centering
\leavevmode
\epsfxsize=0.9
\columnwidth
\epsfbox{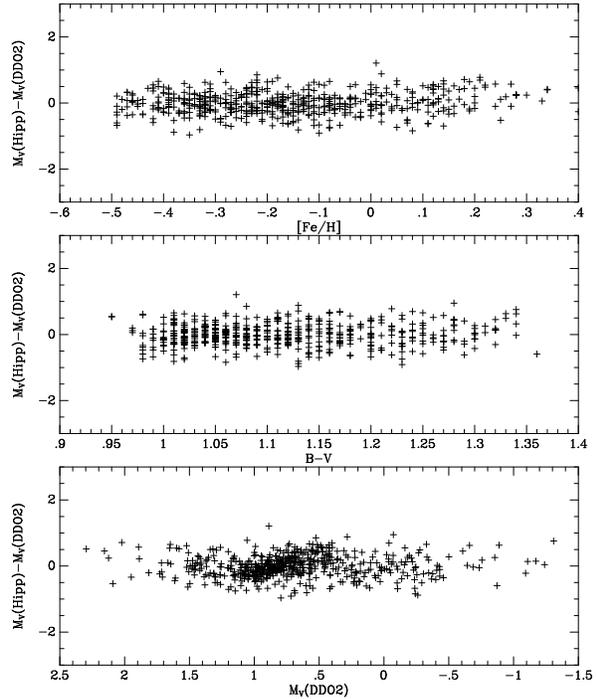}
\vskip 1 cm
\caption{Comparison of the differences between Hipparcos absolute magnitude and
those derived using the new calibration of the DDO system (DDO2).  Upper panel:
as a function of metal abundance [Fe/H]. Middle panel: as a function of colour,
$B-V$. Lower panel: as a function of $M_V(\mathrm{DDO2})$. No significant
residual trends are apparent.  }
\end{figure}

\section{Checking the calibration using K giants in open clusters}

  Our new DDO calibration (DDO2) is based on the Hipparcos absolute magnitudes
for local K giants. Younger giants will therefore be overrepresented in the
calibration sample since they remain closer to the plane than older giants. To
check that this age bias has not led to systematic effects for older giants, we
can check the calibration by examining K giants with DDO photometry in old open
clusters, for whch ages are known. The cluster distance moduli, if based on
main sequence fitting, allow the absolute magnitudes of the giants to be found
independently.  This provides a general check on the calibration and so is
useful in any case.

\begin{table*}
\small
\caption{Open clusters used to check the absolute magnitude calibration}
\begin{tabular}{lrccrrrr}
\hline
Name          &$(M-m)_V$ & Source$^a$ & log(Age)& [Fe/H]   & $E(B-V)$&  $N$  &$\Delta M_V$\\
IC 4651       & 10.32~~~~& 1&    9.28 &$    0.10$&$   0.14~~~~ $&   9  &$    0.119 $\\
Melotte 66    & 12.69~~~~& 2&    9.69 &$   -0.51$&$   0.13~~~~ $&   2  &$   -0.110 $\\ 
NGC ~188      & 11.71~~~~& 1&    9.82 &$   -0.05$&$   0.12~~~~ $&   9  &$    0.201 $\\
NGC ~752      &  7.94~~~~& 1&    9.25 &$   -0.25$&$   0.02~~~~ $&   4  &$   -0.165 $\\ 
NGC 2287      &  9.18~~~~& 1&    8.38 &$   -0.06$&$   0.01~~~~ $&   5  &$    0.356 $\\ 
NGC 2360      & 10.52~~~~& 2&    9.27 &$   -0.28$&$   0.08~~~~ $&   3  &$    0.343 $\\ 
NGC 2420      & 11.85~~~~& 2&    9.47 &$   -0.42$&$   0.02~~~~ $&   5  &$   -0.141 $\\ 
NGC 2506      & 12.60~~~~& 2&    9.36 &$   -0.52$&$   0.02~~~~ $&   3  &$    0.395 $\\ 
NGC 2548      &  9.12~~~~& 3&    8.50 &$   -0.13$&$   0.04~~~~ $&   3  &$   -0.195 $\\ 
NGC 2682 (M67)&  9.58~~~~& 2&    9.70 &$   -0.09$&$   0.05~~~~ $&  25  &$    0.136 $\\ 
NGC 3114      &  9.99~~~~& 4&    7.89 &$   -0.13$&$   0.04~~~~ $&   9  &$   -0.461 $\\ 
NGC 3532      &  8.47~~~~& 1&    8.50 &$   -0.15$&$   0.04~~~~ $&   3  &$   -0.192 $\\ 
NGC 3680      & 10.33~~~~& 2&    9.60 &$   -0.16$&$   0.06~~~~ $&   9  &$    0.735 $\\ 
NGC 3960      & 12.03~~~~& 2&    9.03 &$   -0.34$&$   0.30~~~~ $&   2  &$    0.033 $\\ 
NGC 5822      &  9.68~~~~& 2&    9.08 &$   -0.21$&$   0.12~~~~ $&   6  &$    0.127 $\\ 
NGC 6791      & 13.60~~~~& 5&   10.0~ &$    0.20$&$   0.20~~~~ $&   7  &$   -0.697 $\\ 
NGC 7789      & 12.04~~~~& 2&    9.26 &$   -0.24$&$   0.24~~~~ $&   9  &$   -0.276 $\\ 
\hline
\end{tabular}
\begin{flushleft}
~~~~~~~~$^a$ Sources for apparent distance moduli, $(M-m)_V$ \\
~~~~~~~~1: Meynet, Mermilliod, Maeder (1993)\\
~~~~~~~~2: Friel (1995) \\
~~~~~~~~3: Janes, Tilley, Lyng\aa~(1988) \\
~~~~~~~~4: Sagar, Sharples (1991) \\
~~~~~~~~5: Tripicco, Bell, Dorman, Hufnagel (1995) \\
\end{flushleft}
\end{table*}

  We choose almost the same set of clusters as analysed by Flynn and Mermilliod
(1991). These authors checked the absolute magnitude zero point of the DDO
system, confirming the Janes (1979) analysis, by using much improved distance
estimates (based on main sequence photometry) to open clusters than were
avaliable to Janes in 1979.  Distance estimates to these clusters have improved
still further since 1991, and we list the clusters and the adopted parameters
in Table 2. Two clusters, NGC 2423 and NGC 2099 have been dropped, since the
distance moduli are poorly known relative to the others.  A new cluster, NGC
6791, has been added, since DDO data are now available for it. In Table 2 we
show the cluster name (column 1), the adopted distance modulus and its source
(columns 2 and 3), the log of the age in years, (column 4, log($T$)), the
adopted reddening $E(B-V)$ in column 6 and the number of giants $N$ with DDO
photometry in column 7. The data for these clusters were obtained from the
General Catalogue of Photometric Data maintained by J.-C.~Mermilliod, B.~Hauck
and M.~Mermilliod. Their website (http://obswww.unige.ch/gcpd/gcpd.html)
enabled us to search for all available DDO data in open clusters. We obtained
DDO and broadband $V$ and $B-V$ photometry for the cluster giants from this
site. We de-reddened the DDO photometry using the Janes (1975) method, using
the mean cluster reddening rather than individually determined reddenings, as
this was judged a more stable process. Cluster giants with de-reddened colours
in the range $0.85 < {\mathrm{C}}4245_0 < 1.15$ were then selected (the colour
range of the Hipparcos sample) and this is the number of giants $N$ seen in
column 7 of Table 2. For each giant the absolute magnitude was calculated from
the apparent magnitude and the apparent distance modulus, which we denote
$M_V(\mathrm{distmod})$, and was also calculated from the de-reddened DDO
photometry using the new calibration in the previous section and this we denote
$M_V(\mathrm{DDO2})$. For the giants in each cluster the mean difference
between these was calculated (and shown in column 8 of Table 2)

\begin{figure} 
\input epsf
\centering
\leavevmode
\epsfxsize=0.9
\columnwidth
\epsfbox{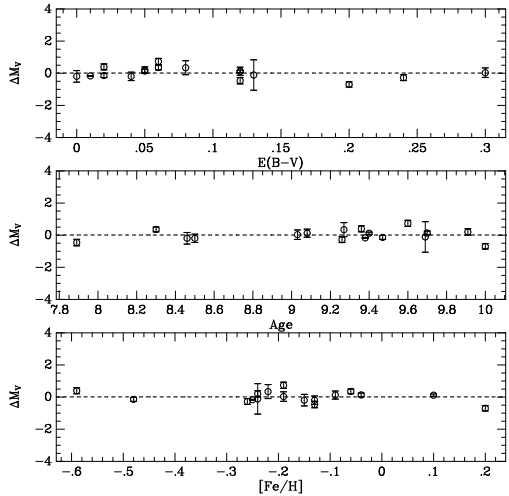}
\vskip -2 cm
\caption{Mean difference $\Delta M_V$ between the DDO derived absolute
magnitudes and the distance modulus derived absolute magnitudes for K
giants in 17 open clusters from Table 2. Upper panel: $\Delta M_V$ as
a function of reddening $E(B-V$). Middle panel: $\Delta M_V$ as a
function of cluster age (in log years), lower panel: $\Delta M_V$ as a
function of cluster abundance [Fe/H]. No significant systematic
residuals are seen as a function of any of these quantities.}
\end{figure}

\begin{equation}
\Delta M_V = M_V(\mathrm{DDO2}) - M_V(\mathrm{distmod}).
\end{equation}

  For the 17 clusters the mean difference is $0.01 \pm 0.10$, with a scatter
amongst the cluster zero points of 0.4 dex.  These results are shown in Figure
7 where for each of the 17 clusters the mean difference in the methods is shown
as a function of cluster abundance [Fe/H], age and reddening.  One can see that
there are no significant trends with age, abundance or reddening. It is
satisfying to see no significant trend with reddening or abundance, although
these are only self-consistency checks on the method. Most important in this
plot is that there is no significant trend with age. Since the calibration was
derived from local K giants, amongst which young K giants will be
overrepresented, there was a potential for a systematic difference in the older
clusters, but this is not seen.  We conclude that the method can be applied to
disc K giants over a full range of age, and further that there are no
significant systematic differences between our absolute magnitude calibration
and the absolute magnitudes of giants in well studied open clusters.  We expect
an improvement in the near future in the distance moduli to some of these
clusters, since Hipparcos has made direct parallax measurements for a large
number of stars in open clusters.

\section{Discussion and Conclusions}

  We have used very accurate Hipparcos data for a sample of 581 local K giants
to measure their absolute magnitudes. These stars were used to check the
absolute magnitudes derived using intermediate band DDO photometry. A number of
systematic offsets in the DDO system emerged from this comparison.  The
Hipparcos data were then used to derive a new calibration of absolute magnitude
in the DDO system. We have checked our new calibration satisfactorily against K
giants with DDO photometry in 17 open clusters. Our calibration is appropriate
for K giants in the colour range $0.85 < $C4245$ < 1.15$ or approximately $0.95
< B-V < 1.3$, and with [Fe/H]$ > -0.5$, (i.e.  metal rich stars) with an
accuracy of 0.35 mag.

  The quality of the Hipparcos data enable a great improvement to be made in
the estimation of K giant $M_V$ by photometric methods. With the release of the
full data set in mid-1997, it should be possible to extend the calibration
presented here to giants with [Fe/H] $< -0.5$ and to subgiants.  K giants have
a venerable tradition in Galactic structure studies, such as tracing inner and
outer disc kinematics, measuring the disc scale length, properties of the
Bulge, and the disc dark matter problem.  The latter was our primary motivation
for this study, since the work of Bahcall, Flynn and Gould (1992), Flynn and
Fuchs (1994) used K giants to trace the scale height and kinematics of the disc
and place limits on its dark matter content. This new calibration will provide
an improved measure of the distances of K giants and we are looking forward to
using it to reanalyse the disc dark matter problem.

\section*{Acknowledgments}

  We thank Jean-Claude Mermilliod for much help and advice with the data in the
open clusters.  This research has made extensive use of the SIMBAD database,
operated at C.D.S., Strasbourg, France, for which we are very grateful. Our
study was supported by the Danish Space Board.
 
{} 

\end{document}